\documentclass[letterpaper,titlepage,11pt]{article}

\usepackage{amssymb,amsmath,amsfonts}
\usepackage{graphicx}
\usepackage{epsfig}
\usepackage{ccaption}
\usepackage{wrapfig}
\usepackage[small,bf,up]{caption}

\usepackage[
      colorlinks=true,
      linkcolor=black,
      urlcolor=black,
      filecolor=black,
      citecolor=black,
      pdfstartview=FitV,
      pdftitle={},
        pdfauthor={Kentaro Tanabe},
        pdfsubject={},
        pdfkeywords={},
        pdfpagemode=None,
        bookmarksopen=true
      ]{hyperref}

\def\clock{{\count0=\time
           \divide\count0 60
           \ifnum\count0<10 0\fi\the\count0
           \multiply\count0 -60 \advance\count0 \time
           :\ifnum\count0<10 0\fi \the\count0
         }}
\newcommand{\timestamp}{{\small\vbox{\hbox{\tt\jobname.tex}
\hbox{\the\day/\the\month/\the\year, \clock}}}}

\newcommand{\sR}{\mathsf{R}}

\newcommand{\tH}{\text{H}}

\newcommand{\hk}{\hat{\kappa}}

\newcommand{\mP}{m_{\Phi}}

\newcommand{\pz}{\partial_{z}}
\newcommand{\pv}{\partial_{v}}
\newcommand{\pp}{\partial_{\phi}}
\newcommand{\pP}{\partial_{\Phi}}

\numberwithin{equation}{section}

\begin{document}

\begin{titlepage}
\rightline{RUP-16-15} 
\leftline{}
\vskip 2cm
\centerline{\LARGE \bf Elastic instability of black rings at large $D$}
\vskip 1.6cm
\centerline{\bf Kentaro Tanabe}
\vskip 0.5cm
\centerline{\sl Department of Physics, Rikkyo University,}
\centerline{\sl  Toshima, Tokyo 171-8501, Japan}
\smallskip
\vskip 0.5cm
\centerline{\small\tt ktanabe@rikkyo.ac.jp}

\vskip 1.6cm
\centerline{\bf Abstract} \vskip 0.2cm \noindent
Using the inverse dimensional expansion method we study the elastic instability of black rings found recently in numerical analysis of fully nonlinear dynamical evolutions. In our analysis we should perform $1/\sqrt{D}$ expansions, not usual $1/D$ expansions,  of the Einstein equations to capture this elastic instability of $D$ dimensional black rings. By solving the Einstein equations at large $D$ we obtain the effective equations for black rings, and the perturbation analysis of the large $D$ effective equations with $1/\sqrt{D}$ expansions yields the formula for quasinormal mode frequencies. From this formula, we find that black rings actually suffer from both elastic and Gregory-Laflamme like instabilities. These instabilities are coupled and appear at the same time as observed in numerical analysis. The elastic instability does disappear at the infinite limit of a ring radius, which implies that the (boosted) black string is stable to elastic perturbations. Furthermore we observe that the Gregory-Laflamme like instability becomes more efficient than the elastic instability even for certain thin black rings in enough higher dimensions.

\end{titlepage}
\pagestyle{empty}
\small
\normalsize
\newpage
\pagestyle{plain}
\setcounter{page}{1}

\section{Introduction}

Black rings \cite{Emparan:2001wn} have various kinds of instabilities. One is the Gregory-Laflamme (GL) like non-axisymmetric instability\footnote{In this paper we consider black rings with single angular momentum in the $S^{1}$ direction. The axisymmetry is the symmetry along this rotating direction. }, which connects with the GL instability of the (boosted) black strings \cite{Gregory:1993vy, Hovdebo:2006jy} at the limit of infinite ring radius of the black ring. Another is the axisymmetric instability in radial perturbations of fat black rings \cite{Elvang:2006dd, Figueras:2011he}. These instabilities of black rings can coexist and cover whole parameter region of the black ring in five dimensions \cite{Santos:2015iua}. So there is no dynamically stable five dimensional black ring. Recently a new type non-axisymmetric instability of five dimensional black rings was found in numerical analysis, so called elastic instability \cite{Figueras:2015hkb}. According to their results, certain thin black rings suffer from not only GL like instability but also this elastic instability, and the elastic instability dominates the evolutions of such black rings over the GL like instability. Then the final state of elastically unstable black rings is a spherical rotating black hole (Myers-Perry black hole) without a violation of the cosmic censorship conjecture. Thus the elastic instability is more important dynamics for such certain thin black rings than the GL like instability. However there are only numerical results and no analytic investigations on it. This is contrasted with the fact that there are analytic or semi-analytic results of the GL like instability and radial perturbation instability \cite{Elvang:2006dd, Figueras:2011he, Tanabe:2015hda}. Hence it is interesting to see if we can capture the elastic instability in analytic way to understand the nature and dimensional dependence of the elastic instabilities. In this paper we use the large $D$ expansion method \cite{Asnin:2007rw, Emparan:2013moa, Emparan:2013xia} to this end. $D$ is a spacetime dimension.  

The large $D$ expansion method has been applied to various black hole physics such as instability of (rotating) black holes and black branes by quasinormal modes \cite{Emparan:2014jca, Emparan:2014aba, Emparan:2015rva, Chen:2015fuf}, constructions of black hole solutions with or without time dependences via effective equations \cite{Tanabe:2015hda, Emparan:2015hwa, Bhattacharyya:2015dva, Suzuki:2015iha, Suzuki:2015axa, Emparan:2015gva} and investigations of charged black holes \cite{Guo:2015swu, Andrade:2015hpa, Bhattacharyya:2015fdk, Tanabe:2015isb, Emparan:2016sjk, Tanabe:2016}.  Especially we obtained $1/D$ expanded black ring solution analytically by solving large $D$ effective equations for slowly rotating large $D$ black holes in our previous analysis \cite{Tanabe:2015hda}. The perturbation analysis of the effective equations gives the quasinormal mode formula, which describes the GL like instability of black rings. However we could not find any elastic instabilities in that formula. This is because of the limitation in our previous analysis. In the paper \cite{Tanabe:2015hda} the azimuthal number, $m_{\Phi}$, along the rotating direction $\Phi$ is assumed to $O(\sqrt{D})$. This assumption is needed to expand the Einstein equations only by $1/D$ since the horizon angular velocity of the black ring has an additional $\sqrt{D}$ factor as $\Omega_{\tH}\sim g_{t\Phi}= O(1/\sqrt{D})$ \cite{Emparan:2007wm}. On the other hand the elastic instability has been found in perturbations with $m_{\Phi}=O(1)$ \cite{Figueras:2015hkb}. To study the physics of perturbations with $m_{\Phi}=O(1)$, we should consider $1/\sqrt{D}$ expansions of the Einstein equations instead of $1/D$ expansions. The purpose of this paper is to give quasinormal mode formula for the modes with $m_{\Phi}=O(1)$ which are expected to describe the elastic instability by performing $1/\sqrt{D}$ expansions of the Einstein equations. 

The organization of this paper is as follows. In section \ref{2} we give a short review on the large $D$ effective equations for black rings. In section \ref{3} we analyze the effective equations by using $1/\sqrt{D}$ expansion, and the elastic instability will be found together with the GL like instability. In section \ref{4} we give the summary of this paper.  Appendixes contain some technical details: ring coordinate embedding in Appendix \ref{A} and quasinormal modes of the boosted black string in Appendix \ref{B}. A {\it Mathematica} file is attached to this paper, which gives the effective equations for slowly rotating large $D$ black holes up to $O(1/D^{2})$ corrections. 

Note that we use the large parameter $n$ defined by
%
\begin{eqnarray}
n=D-4,
\end{eqnarray}
%
instead of $D$. This definition of $n$ is same with the notation in the papers on black rings \cite{Emparan:2007wm, Armas:2014bia}.

\section{Large $D$ effective theory of black rings}
\label{2}

In this section we give a brief review of our previous analysis \cite{Tanabe:2015hda} where we solved the Einstein equations for slowly rotating large $D$ black holes. The slowly rotating black hole contains the black ring, slowly rotating Myers-Perry black hole, slowly boosted black string as solutions of effective equations. The effective equations can be obtained by solving the Einstein equations. The metric ansatz in Eddington-Finkelstein coordinates\footnote{ Here we use $v$ as a time coordinate in the system, while $t$ was used in \cite{Emparan:2015gva, Emparan:2016sjk}. } to solve the Einstein equations is
%
\begin{eqnarray}
&&
ds^{2} = - A dv^{2} -2(u_{v}dv+u_{a}dX^{a})dr \notag \\
&&~~~~~~~~~~~~~
+r^{2}G_{ab}dX^{a}dX^{b}
-2C_{a}dvdX^{a} +r^{2}H^{2} d\Omega^{2}_{n},
\label{mansatz}
\end{eqnarray}
%
where $X^{a}=(z,\Phi)$. $A$, $u_{v}$, $G_{ab}$ and $C_{a}$ are functions of $(v,r,X^{a})$. The shift vector $u_{a}$ is a function of $(v,X^{a})$. $H$ depends only on $z$. These choices of metric functions are taken as gauge choices. The azimuthal coordinate $\Phi$ is the rotation direction of solutions. The slowly rotating large $D$ black holes are defined by the following condition
%
\begin{eqnarray}
g_{v\Phi} = O(1/\sqrt{n}).
\end{eqnarray}
%
This condition is equivalent to $\Omega_{\tH}=O(1/\sqrt{n})$ for stationary black holes, where $\Omega_{\tH}$ is the horizon angular velocity. Then it is useful to introduce the normalized azimuthal coordinate $\phi$ defined by
%
\begin{eqnarray}
\phi = \sqrt{n}\Phi.
\label{Phirel}
\end{eqnarray}
%
Black hole metrics have very large radial gradients as $\partial_{r}=O(n)$ in near zone at large $n$ \cite{Emparan:2013moa}. To examine this large radial gradient feature properly, we use a near zone radial coordinate $\sR$ defined by\footnote{This is one choice of radial coordinates in near zone to realize $\partial_{r}=O(n)$. In \cite{Emparan:2013xia, Emparan:2015hwa, Suzuki:2015iha} we used the different radial coordinate $\rho$, which relates with $\sR$ by $\sR\propto\cosh^{2}{\rho}$.}
%
\begin{eqnarray}
\sR=\left( \frac{r}{r_{0}} \right)^{n},
\end{eqnarray}
%
where $r_{0}$ is a horizon scale, which we set to unity as $r_{0}=1$. Then the leading order of the Einstein equations in $1/n$ expansions become non-linearly coupled ordinary differential equation system with respect to $\partial_{\sR}$. By solving the leading order Einstein equations in coordinates $(v,\sR,z,\phi)$, we obtain the following leading order solution
%
\begin{eqnarray}
&&
ds^{2} = -\left(1-\frac{p_{v}(v,x^{a})}{\sR} \right)dv^{2} +\frac{2H(z)}{\sqrt{1-H'(z)^{2}}}dvdr  \notag \\
&&~~~~~~~~~~~~~~
-\frac{2p_{a}(v,x^{a})}{n\sR}dvdx^{a}
+dz^{2} +\frac{G(z)^{2}d\phi^{2}}{n} +r^{2}H(z)^{2}d\Omega^{2}_{n} +O(n^{-1}),
\label{LOsol}
\end{eqnarray}
%
where $x^{a}=(z,\phi)$. The mass and momentum densities, $p_{v}(v,x^{a})$ and $p_{a}(v,x^{a})$, are introduced as integration functions of $\sR$-integrations. For boundary conditions and detail scaling assumptions of metric functions at large $n$, see \cite{Tanabe:2015hda}. $G(z)$ is a free function of $z$, and $H(z)$ should satisfy the condition 
%
\begin{eqnarray}
1-H'(z)^{2}+H(z)H''(z)=0,
\label{Hcond}
\end{eqnarray}
%
to solve the Einstein equations at the leading order. This condition on $H(z)$ is integrated to
%
\begin{eqnarray}
2\hk = \frac{\sqrt{1-H'(z)^{2}}}{H(z)},
\label{hkdef}
\end{eqnarray}
%
where $\hk$ is an integration constant. $\hk$ gives a size of the surface gravity of stationary solutions as we will see later. $G(z)$ and $H(z)$ are determined if we consider an explicit embedding of the solution into a background spacetime. 

At the next-to-leading order of the Einstein equations in $1/n$ expansions we obtain non-trivial conditions for $p_{v}(v,x^{a})$ and $p_{a}(v,x^{a})$ as
%
\begin{eqnarray}
\pv p_{v} -\frac{H'(z)}{2\hat{\kappa}H(z)}\pz p_{v} -\frac{\pp^{2} p_{v}}{2\hat{\kappa}G(z)^{2}} 
+\frac{\pp p_{\phi}}{G(z)^{2}}+\frac{H'(z)}{H(z)}p_{z}=0,
\label{Deq1}
\end{eqnarray}
%
%
\begin{eqnarray}
&&
\pv p_{\phi} -\frac{H'(z)}{2\hat{\kappa}H(z)}\pz p_{\phi} -\frac{\pp^{2}p_{\phi}}{2\hat{\kappa}G(z)^{2}} 
+\frac{1}{G(z)^{2}}\pp \Biggl[ \frac{p_{\phi}^{2}}{p_{v}} \Biggr] \notag \\
&&~~~~~
-\frac{4\hat{\kappa}^{2}G(z)H(z)^{2}+2G'(z)H(z)H'(z)}{4\hat{\kappa}^{2}G(z)H(z)^{2}}\pp p_{v}
 \notag \\
&&~~~~~
+\frac{H'(z)}{H(z)}\frac{p_{z}p_{\phi}}{p_{v}}
+\frac{G'(z)H'(z)}{\hat{\kappa}G(z)H(z)}p_{\phi}=0,
\label{Deq2}
\end{eqnarray}
%
and
%
\begin{eqnarray}
&&
\pv p_{z} -\frac{H'(z)}{2\hat{\kappa}H(z)}\pz p_{z} -\frac{\pp^{2}p_{z}}{2\hat{\kappa}G(z)^{2}}
+\pz p_{v} +\frac{1}{G(z)^{2}}\pp \Biggl[ \frac{p_{\phi}p_{z}}{p_{v}} \Biggr] \notag \\
&&~~~~~~~
+\frac{H'(z)}{H(z)}\frac{p^{2}_{z}}{p_{v}}
-\frac{G'(z)}{G(z)^{3}}\frac{p^{2}_{\phi}}{p_{v}} +\frac{G'(z)}{\hat{\kappa}G(z)^{3}}\pp p_{\phi} \notag \\
&&~~~~~~~
+\frac{H(z)G'(z)^{2}H'(z)+G(z)(G'(z)-H(z)H'(z)G''(z))}{4\hat{\kappa}^{2}G(z)^{2}H(z)^{2}}p_{v} \notag \\
&&~~~~~~~
-\frac{1-2H'(z)^{2}}{2\hat{\kappa}H(z)^{2}}p_{z}=0.
\label{Deq3}
\end{eqnarray}
%
These are large $D$ effective equations for general slowly rotating large $D$ black holes. The effective equations are equations for local quantities, $p_{v}(v,x^{a})$ and $p_{a}(v,x^{a})$. The whole shape of a solution such as its topology is determined by the embedding functions $G(z)$ and $H(z)$. For example, if we embed the solution into a flat spacetime in spherical coordinates by proper choice for embedding functions, a solution of effective equations represents an asymptotically flat black hole with spherical horizon topology. Now, by embedding solutions into a flat spacetime in the ring coordinate, we obtain the effective equations for black rings\footnote{As discussed in \cite{Tanabe:2015hda, Emparan:2013moa}, black rings are expected to have $O(1/\sqrt{n})$ horizon angular velocity at large $D$ in general. So we do not need to restrict to "slowly rotating black rings", and effective equations (\ref{Deq1}), (\ref{Deq2}) and (\ref{Deq3}) describe dynamics of general black rings at large $D$ except for extremely fat black rings. }. The ring coordinate embedding is given by
%
\begin{eqnarray}
H(z)=\frac{R  \sin{\theta}}{R+\cos{\theta}},~~
G(z)=\frac{R \sqrt{R^{2}-1}}{R+\cos{\theta}},~~
\frac{d\theta}{dz}=\frac{R+\cos{\theta}}{R }.
\label{BR}
\end{eqnarray}
%
This $H(z)$ satisfies the condition for $H(z)$ given in eq. (\ref{Hcond}). For the detail of the ring coordinate embedding, see Appendix \ref{A}. 
$R$ is a ring radius satisfying $R>1$. Then the effective equations in this ring coordinate embedding become
%
\begin{eqnarray}
&&
\pv p_{v} +\frac{(R+y)(1+Ry)}{R\sqrt{R^{2}-1}}\partial_{y}p_{v}-\frac{(R+y)^{2}}{R(R^{2}-1)^{3/2}}\pp^{2}p_{v} \notag\\
&&~~~~~~~~~~~~~~~~~~~~~
+\frac{(R+y)^{2}}{R^{2}(R^{2}-1)}\pp p_{\phi} +\frac{1+Ry}{R\sqrt{1-y^{2}}}p_{z}=0,
\label{Deq1BR}
\end{eqnarray}
%
%
\begin{eqnarray}
&&
\pv p_{\phi} +\frac{(R+y)(1+Ry)}{R\sqrt{R^{2}-1}}\partial_{y}p_{\phi}-\frac{(R+y)^{2}}{R(R^{2}-1)^{3/2}}\pp^{2}p_{\phi} \notag\\
&&~~~~~~~~
+\frac{(R+y)^{2}}{R^{2}(R^{2}-1)}\pp\Biggl[\frac{p_{\phi}^{2} }{p_{v}}\Biggr]-\frac{1+2Ry+R^{2}}{R^{2}-1}\pp p_{v} \notag \\
&&~~~~~~~~
+\frac{1+Ry}{R\sqrt{1-y^{2}}}\frac{p_{z}p_{\phi}}{p_{v}}+\frac{2(1+Ry)}{R\sqrt{R^{2}-1}}p_{\phi}=0
\label{Deq2BR}
\end{eqnarray}
%
and
%
\begin{eqnarray}
&&
\pv p_{z} +\frac{(R+y)(1+Ry)}{R\sqrt{R^{2}-1}}\partial_{y}p_{z}-\frac{(R+y)^{2}}{R(R^{2}-1)^{3/2}}\pp^{2}p_{z} \notag\\
&&~~~~~~~~
-\frac{(R+y)\sqrt{1-y^{2}}}{R}\partial_{y} p_{v} +\frac{(R+y)^{2}}{R^{2}(R^{2}-1)}\pp\Biggl[ \frac{p_{z}p_{\phi}}{p_{v}} \Biggr]
+\frac{1+Ry}{R\sqrt{1-y^{2}}}\frac{p_{z}^{2}}{p_{v}} \notag \\
&&~~~~~~~~
-\frac{(R+y)^{2}\sqrt{1-y^{2}}}{R^{3}(R^{2}-1)}\frac{p_{\phi}^{2}}{p_{v}}
+\frac{2(R+y)^{2}\sqrt{1-y^{2}}}{R^{2}(R^{2}-1)^{3/2}}\pp p_{\phi}
+\frac{(R+y)\sqrt{1-y^{2}}}{R^{2}-1}p_{v} \notag \\
&&~~~~~~~~
+\frac{2+2Ry-y^{2}+R^{2}(2y^{2}-1)}{R\sqrt{R^{2}-1}(1-y^{2})}p_{z}=0.
\label{Deq3BR}
\end{eqnarray}
%
Here we introduced a coordinate $y$ defined by
%
\begin{eqnarray}
y=\cos{\theta}.
\end{eqnarray}
%
The effective equations (\ref{Deq1BR}), (\ref{Deq2BR}) and (\ref{Deq3BR}) in the ring coordinate embedding  have a black ring solution as a stationary solution, $p_{v}=p_{v}(y)$ and $p_{a}=p_{a}(y)$, given by
%
\begin{eqnarray}
p_{z}(y) = -\frac{(R+y)\sqrt{1-y^{2}}}{\sqrt{R^{2}-1}}p_{v}'(y),~~
p_{\phi} = \hat{\Omega}_{H}\frac{R^{2}(R^{2}-1)}{(R+y)^{2}}p_{v}(y), 
\end{eqnarray}
%
and $p_{v}=e^{P(y)}$ where
%
\begin{eqnarray}
P(y) = p_{0} +\frac{d_{0}}{R+y} +\frac{(1+Ry)(1+Ry+2R(R+y)\log{(R+y)})}{2R^{2}(R+y)^{2}}.
\label{Psol}
\end{eqnarray}
%
$\hat{\Omega}_{\tH}$, $p_{0}$ and $d_{0}$ are integration constants. $\hat{\Omega}_{\tH}$ gives a size of the horizon angular velocity. $p_{0}$ and $d_{0}$ are trivial deformation parameters of the solution such as $O(1/n)$ redefinitions of $R$ and $r_{0}$. We set to $p_{0}=0$ and $d_{0}=0$. In the derivation of eq. (\ref{Psol}), $\hat{\Omega}_{\tH}$ is found to be
%
\begin{eqnarray}
\hat{\Omega}_{\tH} = \frac{\sqrt{R^{2}-1}}{R^{2}},
\end{eqnarray}
%
to have regular solutions at $y=-1/R$. This implies that the horizon angular velocity should be chosen so that the horizon geometry is regular. This is similar situation with five and higher dimensional black ring \cite{Emparan:2001wn,Emparan:2007wm} whose horizon angular velocity is determined by the balance condition between the centrifugal force and tension of the horizon. The stationary solution has a Killing vector
%
\begin{eqnarray}
\partial_{v} +\hat{\Omega}_{\tH}\pp = \partial_{v} +\Omega_{\tH}\pP, 
\end{eqnarray}
%
where $\Omega_{\tH}=\hat{\Omega}_{\tH}/\sqrt{n}$ is the horizon angular velocity. This Killing vector gives a surface gravity $\kappa$ of the stationary solution by
%
\begin{eqnarray}
\kappa = n\hk = n\frac{\sqrt{R^{2}-1}}{2R},
\label{kappa}
\end{eqnarray}
%
at the leading order in $1/n$ expansion. Here we used eqs. (\ref{hkdef}) and (\ref{BR}). The effective equations (\ref{Deq1BR}), (\ref{Deq2BR}) and (\ref{Deq3BR}) describe the non-linear dynamical evolutions of black rings. However we restrict our attention to the linear analysis of the effective equations in this paper.

\paragraph{Quasinormal modes}

Let us consider perturbations of effective equations (\ref{Deq1BR}), (\ref{Deq2BR}) and (\ref{Deq3BR}) around the black ring solution to obtain quasinormal mode formula. The perturbation ansatz is
%
\begin{eqnarray}
&&
p_{v}(v,y,\phi) = e^{P(y)} \left( 1+ \epsilon e^{-i\omega v}e^{i m\phi}F_{v}(y) \right), \\
&&
p_{z}(v,y,\phi)=e^{P(y)}\left( -\frac{(R+y)\sqrt{1-y^{2}}}{\sqrt{R^{2}-1}}P'(y) + \epsilon e^{-i\omega v}e^{i m\phi}F_{z}(y) \right), \\
&&
p_{\phi}(v,y,\phi)= e^{P(y)}\left( \frac{R(R^{2}-1)^{3/2}}{(R+y)^{2}}+ \epsilon e^{-i\omega v}e^{i m\phi}F_{\phi}(y) \right).
\end{eqnarray}
%
$\epsilon$ is a perturbation parameter. $F_{v}(y)$ and $F_{a}(y)$ are perturbation variables. The perturbation equations has a pole at $y=-1/R$. So we should impose regularity conditions on perturbation variables at the pole. The regularity condition at the pole is
%
\begin{eqnarray}
F_{v}(y) \propto (1+Ry)^{\ell}\left(
1+O(1+R y)
\right),
\label{polecond}
\end{eqnarray}
%
where $\ell$ is a positive integer. This regularity condition for perturbation variables at the pole gives a non-trivial condition for the frequency $\omega$. It is the quasinormal mode formula of black rings. The quasinormal mode formula is obtained as \cite{Tanabe:2015hda} 
%
\begin{eqnarray}
&&
\frac{1}{\sqrt{R^{2}-1}(m^{2}+imR+\ell R^{2})-iR^{3}\omega}\Bigl[
R^{9}\omega^{3} + iR^{6}\sqrt{R^{2}-1}\Bigl( 3m^{2}+3imR \notag \\
&&~~~
+(3\ell-2)R^{2} \Bigr)\omega^{2}
-R^{3}(R^{2}-1)\Bigl( 3m^{4}+6im^{3}R+2(3\ell-4) m^{2}R^{2} \notag \\
&&~~~+2i(3\ell-2)m R^{3}+3(\ell-1)\ell R^{4}\Bigr)\omega 
-i(R^{2}-1)^{3/2}\Bigl( m^{6}+3im^{5}R +3(\ell-2)m^{4}R^{2} \notag \\
&&~~~
+6i(\ell-1)m^{3}R^{3} 
-(4-7\ell+3\ell^{2})m^{2}R^{4}+3i\ell(\ell-1)mR^{5} +\ell^{2}(\ell-1)R^{6} \Bigr)
\Bigr] \notag \\
&&~~~
=0.
\label{QNMcondBR}
\end{eqnarray}
%
For example, this formula is solved explicitly for the mode with $\ell=0$, which contains the GL like instability as
%
\begin{eqnarray}
\omega^{(\ell=0)}_{\pm}= \frac{\sqrt{R^{2}-1}}{R}\Bigl[ \hat{m} \pm i\hat{m}(1 \mp \hat{m}) \Bigr],
\label{QNMell0BR}
\end{eqnarray}
%
where $\hat{m}=m/R$. $\omega^{(\ell=0)}_{+}$ is the GL like unstable mode of black rings. For perturbations with $\ell>0$ we found no unstable mode from the formula (\ref{QNMcondBR}). 

There is one important remark on the azimuthal number $m$. $m$ is an azimuthal number along the normalized azimuthal coordinate $\phi$, not physical one $\Phi$. From the relation (\ref{Phirel}), the physical azimuthal number $m_{\Phi}$ along the physical coordinate $\Phi$ is related with $m$ by
%
\begin{eqnarray}
m_{\Phi} =\sqrt{n}m. 
\label{mrel2}
\end{eqnarray}
%
$m_{\Phi}$ is an integer, so $m$ does not need to be an integer in general. $m=O(1)$ implies we consider perturbations with very large physical azimuthal number $m_{\Phi}=O(\sqrt{n})$. The fact there is no unstable mode with $m=O(1)$ and $\ell>0$ suggests that perturbations with $m_{\Phi}=O(\sqrt{n})$ has only one unstable mode, the GL like instability. So there is no elastic instability in the modes with $m_{\Phi}=O(\sqrt{n})$. In next section we study the perturbations with $m_{\Phi}=O(1)$ as considered in \cite{Santos:2015iua, Figueras:2015hkb}.

In the paper \cite{Tanabe:2015hda} we also obtained $O(1/n)$ corrections to the effective equations by solving the Einstein equations up to next-to-next-to-leading order. As one example of this $O(1/n)$ corrections we obtained the quasinormal mode frequency of GL-like unstable modes as
%
\begin{eqnarray}
\omega_{\pm}^{(\ell=0)} = \frac{\sqrt{R^{2}-1}}{R}\Biggl[
\hat{m}\pm i\hat{m}(1\mp\hat{m})+\frac{\delta\hat{\omega}^{(\ell=0)}_{\pm}}{n}
\Biggr],
\end{eqnarray}
%
where
%
\begin{eqnarray}
&&
\delta\hat{\omega}^{(\ell=0)}_{\pm}=\frac{1}{2\hat{m}R^{2}}\Bigl[
\hat{m}^{2}(R^{2}-4+4R^{2}\hat{m}^{2})+i\hat{m}(2+(R^{2}+16)\hat{m}^{2}+8\hat{m}^{4}) \notag \\
&&~~~~~~~~~~~~~~~
+2\hat{m}^{2}(1-2i\hat{m})\log{(R-R^{-1})}
\mp \Bigl(
2\hat{m}(2+(3R^{2}-4)\hat{m}^{2}) \notag \\
&&~~~~~~~~~~~~~~~
+i(2+(3R^{2}+1)\hat{m}^{2}-2(R^{2}-10)\hat{m}^{4}-2\hat{m}^{2}\log{(R-R^{-1})})
\Bigr)
\Bigr].
\label{QNMBRNLOell0}
\end{eqnarray}
%
$\hat{\omega}^{(\ell=0)}_{+}$ gives the GL like instability of black rings up to $O(1/n)$ corrections. The $O(1/n)$ correction $\delta\hat{\omega}^{(\ell=0)}_{\pm}$ contains $m$ in its denominator. So we should be careful to take the small limit of $m$. 
 
To investigate the elastic instability, however, we further need $O(1/n^{2})$ corrections to effective equations. This corrections can be obtained straightforwardly by solving the Einstein equations up to one more higher order. The explicit form of corrections are messy, so we do not show them here. A {\it Mathematica} file attached to this paper contains the detail form of effective equations of slowly rotating large $D$ black holes up to $O(1/n^{2})$ corrections.

\section{Elastic instability}
\label{3}

The elastic instability has been found in the modes with $m_{\Phi}=O(1)$ \cite{Figueras:2015hkb}. So we study perturbations with $m_{\Phi}=O(1)$ of the effective equations (\ref{Deq1BR}), (\ref{Deq2BR}) and (\ref{Deq3BR}). To do this, we rewrite the effective equations by the physical azimuthal coordinate $\Phi$. The relation (\ref{Phirel}) gives
%
\begin{eqnarray}
\pp = \frac{1}{\sqrt{n}}\partial_{\Phi}.
\label{Phirel2}
\end{eqnarray}
%
Hence the effective equations have $1/\sqrt{n}$ series, not $1/n$ series, in terms of the physical coordinate $\Phi$. The stationary solution of effective equations does not have $\phi$ dependences, so the black ring solution obtained in \cite{Tanabe:2015hda}  is not affected by this relation. However the non-axisymmetric perturbation changes its property drastically by this relation as we will see below.

\subsection{Effective equations}

By using eq. (\ref{Phirel2}) we can expand effective equations (\ref{Deq1BR}), (\ref{Deq2BR}) and (\ref{Deq3BR})  by $1/\sqrt{n}$ as\footnote{
One may think that we should resolve the Einstein equations by using $1/\sqrt{n}$ expansions to obtain the effective equations for elastic instabilities. But this is not the case. The $\sqrt{n}$ factor appears only through $\pP$ operators. So, if we already have effective equations with respect to $\pp$, we can obtain effective equations of slowly rotating large $D$ black holes in $1/\sqrt{n}$ expansions immediately just by using eq. (\ref{Phirel2}) to effective equations in $1/n$ expansions. 
}
%
\begin{eqnarray}
&&
\pv p_{v} +\frac{(R+y)(1+Ry)}{R\sqrt{R^{2}-1}}\partial_{y}p_{v}
+\frac{1+Ry}{R\sqrt{1-y^{2}}}p_{z}
+\frac{1}{\sqrt{n}}\frac{(R+y)^{2}}{R^{2}(R^{2}-1)}\pP p_{\phi}=O(n^{-1}),
\label{Deq1BR2}
\end{eqnarray}
%
%
\begin{eqnarray}
&&
\pv p_{\phi} +\frac{(R+y)(1+Ry)}{R\sqrt{R^{2}-1}}\partial_{y}p_{\phi} 
-\frac{1+2Ry+R^{2}}{R^{2}-1}\pp p_{v} +\frac{1+Ry}{R\sqrt{1-y^{2}}}\frac{p_{z}p_{\phi}}{p_{v}}\notag \\
&&~~~~~~~~
+\frac{2(1+Ry)}{R\sqrt{R^{2}-1}}p_{\phi}
+\frac{1}{\sqrt{n}}\frac{(R+y)^{2}}{R^{2}(R^{2}-1)}\pP\Biggl[\frac{p_{\phi}^{2} }{p_{v}}\Biggr]=O(n^{-1})
\label{Deq2BR2}
\end{eqnarray}
%
and
%
\begin{eqnarray}
&&
\pv p_{z} +\frac{(R+y)(1+Ry)}{R\sqrt{R^{2}-1}}\partial_{y}p_{z}
-\frac{(R+y)\sqrt{1-y^{2}}}{R}\partial_{y} p_{v} +\frac{1+Ry}{R\sqrt{1-y^{2}}}\frac{p_{z}^{2}}{p_{v}} \notag \\
&&~~~~~~~~
-\frac{(R+y)^{2}\sqrt{1-y^{2}}}{R^{3}(R^{2}-1)}\frac{p_{\phi}^{2}}{p_{v}}
+\frac{(R+y)\sqrt{1-y^{2}}}{R^{2}-1}p_{v} 
+\frac{2+2Ry-y^{2}+R^{2}(2y^{2}-1)}{R\sqrt{R^{2}-1}(1-y^{2})}p_{z}
 \notag \\
&&~~~~~~~~
+\frac{1}{\sqrt{n}}\Biggl[
\frac{(R+y)^{2}}{R^{2}(R^{2}-1)}\pP\Biggl[ \frac{p_{z}p_{\phi}}{p_{v}} \Biggr]
+\frac{2(R+y)^{2}\sqrt{1-y^{2}}}{R^{2}(R^{2}-1)^{3/2}}\pP p_{\phi}
\Biggr]
=O(n^{-1}).
\label{Deq3BR2}
\end{eqnarray}
%
Now we have effective equations up to $O(1/n^{2})$ corrections in terms of $\phi$ coordinate. This means that we have $O(1/n^{5/2})$ corrections in $\Phi$ coordinate. Here, however, we show up to only $O(1/\sqrt{n})$ corrections for simplicity of representations. The higher order corrections can be obtained by similar manner with the usage of eq. (\ref{Phirel2}). 

\paragraph{Perturbation equations}

The black ring solution is the stationary solution of eq. (\ref{Deq1BR2}), (\ref{Deq2BR2}) and (\ref{Deq3BR2}). In the paper \cite{Tanabe:2015hda} the black ring solution was obtained up to $O(1/n)$ corrections as
\footnote{The explicit solution for $p_{z}$ has very lengthy form, so we do not show it.}
%
\begin{eqnarray}
p_{v}(y)=e^{P(y)}\Biggl[ 1+\frac{P^{(1)}(y)}{n} \Biggr],
\end{eqnarray}
%
%
\begin{eqnarray}
&&
p_{\phi}(y)=\frac{R(R^{2}-1)^{3/2}}{(R+y)^{2}}e^{P(y)} \notag \\
&&~~~\times
\Biggl[ 1+
\frac{1}{n} \Bigl(
P^{(1)}(y) -\frac{2(1+Ry)}{R^{2}-1}P(y)
+\frac{\log{(R-R^{-1})}}{R^{2}}
\Bigr)
\Biggr],
\end{eqnarray}
%
where $P^{(1)}(y)$ is given by
%
\begin{eqnarray}
&&
P^{(1)}(y) = p_{1}+\frac{d_{1}}{R+y} +\frac{(R^{2}-1)^{2}}{R^{4}(R+y)^{2}}\log{(R-R^{-1})}\notag \\
&&~~~~~
-\frac{2(1+Ry)}{R^{3}(R+y)}\text{Li}_{2}\left( \frac{1+Ry}{R^{2}-1} \right) 
+\frac{(R^{2}-1)(y-R(1-2y^{2}))}{2R^{3}(R+y)^{2}}(\log{(R+y)})^{2}\notag \\
&&~~~~~
-\frac{1}{4R^{4}(R^{2}-1)(R+y)^{4}}\Bigl[
3+12Ry+2y^{2}+2R^{8}(4-3y^{2})+2R^{7}y(5-4y^{2}) \notag \\
&&~~~~~~~~~
-R^{2}(5-6y^{2})-12R^{5}y(1+y^{2})
-R^{6}(12-8y^{2}+5y^{4}) 
-2R^{3}y(10+y^{4})\notag \\
&&~~~~~~~~~~
+R^{4}(4-30y^{2}-5y^{4})
\Bigr] \notag \\
&&~~~~~
+\frac{\log{(R+y)}}{R^{3}(R^{2}-1)(R+y)^{3}}\Bigl[
-2+R^{7}y-y^{2}+3R^{2}(1-2y^{2})+R^{6}(2+y^{2})\notag \\
&&~~~~~~~~~~
+3R^{3}y(2+y^{2})-2Ry(3+y^{2}) 
+3R^{5}y(1+y^{2}) 
-R^{4}(2-12y^{2}+y^{4})\notag \\
&&~~~~~~~~~~
+2(1+Ry)(R^{2}-1)(R+y)^{2}\log{(R-R^{-1})}
\Bigr].
\end{eqnarray}
%
$p_{1}$ and $d_{1}$ are integration constants representing trivial deformation of the solution. $\text{Li}_{k}(x)$ is the polylogarithm function.  Using $O(1/n^{2})$ correction to effective equations we can obtain the black ring solution up to $O(1/n^{2})$ corrections, which we use as the background of the perturbation. We consider the perturbation of this black ring solution as
%
\begin{eqnarray}
&&
p_{v}(v,y,\phi) = e^{P(y)} \left( 1+ \epsilon e^{-i\omega v}e^{i m_{\Phi}\Phi}F_{v}(y) \right), \\
&&
p_{z}(v,y,\phi)=e^{P(y)}\left( -\frac{(R+y)\sqrt{1-y^{2}}}{\sqrt{R^{2}-1}}P'(y) + \epsilon e^{-i\omega v}e^{i m_{\Phi}\Phi}F_{z}(y) \right), \\
&&
p_{\phi}(v,y,\phi)= e^{P(y)}\left( \frac{R(R^{2}-1)^{3/2}}{(R+y)^{2}}+ \epsilon e^{-i\omega v}e^{i m_{\Phi}\Phi}F_{\phi}(y) \right),
\end{eqnarray}
%
where $|m_{\Phi}|=0,1,2,3,..$ The difference from the previous analysis \cite{Tanabe:2015hda} is only the fact that we use $e^{im_{\Phi}\Phi}$, not $e^{im\phi}$ for perturbation decompositions. 
Due to the relation (\ref{Phirel2}), each perturbation variables, $F_{v}(y)$ and $F_{a}(y)$, have following expansions in $1/\sqrt{n}$ 
%
\begin{eqnarray}
F_{v}(y) = \sum_{k\geq 0}\frac{F^{(k)}_{v}(y)}{n^{k/2}},~~
F_{a}(y) = \sum_{k\geq 0}\frac{F^{(k)}_{a}(y)}{n^{k/2}}.
\end{eqnarray}
%
Then perturbation equations at each orders are given by
%
\begin{eqnarray}
\frac{d}{dy}(P(y)F^{(k)}_{v}(y) )-\frac{i\omega R\sqrt{R^{2}-1}}{(R+y)(1+Ry)}F^{(k)}_{v}(y) +\frac{\sqrt{R^{2}-1}}{(R+y)\sqrt{1-y^{2}}}F_{z}^{(k)}(y) = \mathcal{S}^{(k)}_{v},
\label{pDeq1BR}
\end{eqnarray}
%
%
\begin{eqnarray}
&&
\frac{d}{dy}F^{(k)}_{z}(y) -\frac{i\omega R\sqrt{R^{2}-1}}{(R+y)(1+Ry)}F^{(k)}_{z}(y) 
-\left( P'(y) -\frac{2(1+Ry)-(R^{2}-2R^{2}y^{2}+y^{2})}{(1-y^{2})(R+y)(1+Ry)} \right)F_{z}^{(k)}(y) \notag \\
&&~~~~
-\frac{\sqrt{(R^{2}-1)(1-y^{2})}}{1+Ry}\frac{d}{dy}F^{(k)}_{v}(y) 
-\frac{2(R^{2}-1)\sqrt{1-y^{2}}}{R^{2}(R+y)(1+Ry)}F_{\phi}^{(k)}(y) \notag \\
&&~~~~
+\frac{\sqrt{(R^{2}-1)(1-y^{2})}}{R^{2}(R+y)^{3}(1+Ry)}\Biggl[
(R^{2}-1)^{2} +\frac{R^{3}(R+y)^{3}}{R^{2}-1} -R^{2}(R+y)^{3}P'(y) \notag \\
&&~~~~~~~~~~~~~~~~~~~~
-\frac{R^{2}(R+y)^{4}(1+Ry)}{R^{2}-1}P'(y)^{2}
\Biggr]F_{v}^{(k)}(y) 
 =\mathcal{S}^{(k)}_{z},
 \label{pDeq2BR}
\end{eqnarray}
%
and
%
\begin{eqnarray}
&&
\frac{d}{dy}F_{\phi}^{(k)}(y) -\frac{i\omega R\sqrt{R^{2}-1}}{(R+y)(1+Ry)}F^{(k)}_{\phi}(y) +\frac{2}{R+y}F_{\phi}^{(k)}(y) +\frac{(R^{2}-1)^{2}}{(R+y)^{3}\sqrt{1-y^{2}}}F^{(k)}_{z}(y) \notag \\
&&~~~~~~~
+\frac{(R^{2}-1)^{3/2}}{(R+y)^{2}}P'(y)F^{(k)}_{v}(y) =\mathcal{S}_{\phi}^{(k)}.
\label{pDeq3BR}
\end{eqnarray}
%
$\mathcal{S}^{(k)}_{v}$, $\mathcal{S}^{(k)}_{z}$ and $\mathcal{S}^{(k)}_{\phi}$ are source terms for $k$-th order variables, which are functions of lower order perturbation solutions. For example, at $k=1$,  we have
%
\begin{eqnarray}
\mathcal{S}^{(1)}_{v} = \frac{im_{\Phi}}{R\sqrt{R^{2}-1}(1+Ry)}F_{\phi}^{(0)}(y),
\label{Sv}
\end{eqnarray}
%
%
\begin{eqnarray}
\mathcal{S}^{(1)}_{z} = \frac{im_{\Phi}(R^{2}-1)}{R(R+y)(1+Ry)}F_{z}^{(0)}(y)
-\frac{im_{\Phi}\sqrt{1-y^{2}}((R+y)P'(y)-2)}{R(R^{2}-1)(1+Ry)}F_{\phi}^{(0)}(y),
\label{Sz}
\end{eqnarray}
%
and
%
\begin{eqnarray}
&&
\mathcal{S}^{(1)}_{\phi} = \frac{2im_{\Phi}(R^{2}-1)}{R(R+y)(1+Ry)}F_{\phi}^{(0)}(y) \notag \\
&&~~~~~~~
-\frac{im_{\Phi}}{R(R^{2}-1)(R+y)^{3}(1+Ry)}\Bigl[
2R^{6}+4R^{5}y+R^{2}(y^{2}+3) \notag \\
&&~~~~~~~~~~~~~~~~~~~
+R^{4}(5y^{2}-2) +2R^{3}y(1+y^{2})-1
\Bigr]F_{v}^{(0)}(y).
\label{Sp}
\end{eqnarray}
%

\subsection{Quasinormal modes}

As we can see in eqs. (\ref{pDeq1BR}), (\ref{pDeq2BR}) and (\ref{pDeq3BR}) or source terms (\ref{Sv}), (\ref{Sz}) and (\ref{Sp}), the perturbation equations have a pole at $y=-1/R$. So solutions of perturbation equations have singular behaviors at $y=-1/R$ in general. To avoid this singular behavior we impose the condition
%
\begin{eqnarray}
F_{v}^{(k)}(y) \propto (1+Ry)^{\ell}(1+O(1+Ry))
\label{regcond}
\end{eqnarray}
%
with a positive integer $\ell$ on the perturbation variables as regularity conditions. Then we obtain regular perturbation solution at $y=-1/R$. This is same situation with the case of $m_{\Phi}=O(\sqrt{n})$ in previous section. To satisfy the regularity condition (\ref{regcond}), we find that there is a condition on the frequency $\omega$, which is the quasinormal mode formula of the black ring. In the following we give results of the quasinormal mode formula for $\ell>1$ and for $\ell=0,1$ separately. 

\paragraph{$\ell>1$ modes}

For $\ell>1$ modes we obtain the quasinormal mode formula by
%
\begin{eqnarray}
&&
\omega^{(\ell>1)}_{\pm} = \pm\sqrt{\ell-1} -i(\ell-1) +\frac{1}{\sqrt{n}}\frac{m_{\Psi}}{R}  \notag \\
&&~~~~
+\frac{1}{n}\Biggl[
\pm \frac{\ell^{2}R^{2}(6R^{2}-7)-\ell(16R^{4}-R^{2}-16)+10(R^{4}+R^{2}-2)}{4\sqrt{\ell-1}R^{2}(R^{2}-1)} 
\mp \frac{(\ell-2)R^{2}}{2\sqrt{\ell-1}\ell R^{2}} m_{\Phi}^{2} \notag \\
&&~~~~~~~~
-i\frac{4\ell^{2}R^{2} (R^{2}-1) -\ell(14R^{4}-7R^{2}+8)+10(R^{4}+R^{2}-2)}{4R^{2}(R^{2}-1)} -i\frac{\ell-1}{\ell R^{2}} m_{\Phi}^{2}
\Biggr] \notag \\
&&~~~~~~~
+O(n^{-3/2}).
\label{QNMell2}
\end{eqnarray}
%
Here we show only up to $O(1/n)$ corrections. The leading order frequency is same with one of the Schwarzschild black hole \cite{Emparan:2014aba, Emparan:2015rva, Bhattacharyya:2015dva}. So each section at $\Phi=\text{constant}$ surface is the Schwarzschild black hole for the perturbations with $m_{\Phi}=O(1)$ at the leading order of large $D$. The difference from the Schwarzschild black hole appears from $O(1/n)$ in the frequency\footnote{$O(n^{-1/2})$ part is understood just as the boost effect of the black string. }.  As we can see, this formula cannot be applied to $\ell=0,1$ modes since $O(1/n)$ corrections have singular behaviors for the modes. This is because $\ell=0$ and $\ell=1$ modes become coupled, and we cannot treat them separately unlike $\ell>1$ modes. From eq. (\ref{QNMell2}), the perturbations with $\ell>1$ are stable even for $m_{\Phi}=O(1)$\footnote{For $m_{\Phi}=O(\sqrt{n})$ the perturbation with $\ell>0$ are also stable as we observed in \cite{Tanabe:2015hda} and previous section.
}. So, if black rings are unstable, such instability modes should be in $\ell=0$ or $\ell=1$. In fact the GL like instability is in the mode with $\ell=0$, and the elastic instability is in the mode with $\ell=1$.

\paragraph{$\ell=0,1$ modes}
At first one may think that the quasinormal mode frequencies for perturbations with $m_{\Phi}=O(1)$ can be obtained by applying eq. (\ref{mrel2}) to the quasinormal mode condition (\ref{QNMcondBR}), and actually, this is true for $\ell>1$ modes. However it is not the case for $\ell=0$ and $\ell=1$ modes. This can be seen in the quasinormal mode (\ref{QNMBRNLOell0}) for $\ell=0$ mode perturbations. The formula (\ref{QNMBRNLOell0}) is not applicable to perturbations with $m_{\Phi}=m\sqrt{n}=O(1)$ since $O(1/n)$ correction becomes same order with the leading order for $m=O(1/\sqrt{n})$. In fact if we impose the regularity condition (\ref{regcond}) with $\ell=0$ or $\ell=1$, we cannot have non-trivial regular perturbation solution. This can be understood also from the fact that the formula (\ref{QNMell2}) becomes singular for $\ell=0$ and $\ell=1$.  Hence we should impose different regularity conditions from eq. (\ref{regcond}) for the perturbations with $\ell=0,1$. Instead, by allowing both behaviors of $\ell=0$ and $\ell=1$ in eq. (\ref{regcond}) for perturbations, we obtain the non-zero regular perturbation with quasinormal mode formula. Actually $F_{v}^{(0)}$ and $F_{v}^{(1)}$ can satisfy the regularity condition with $\ell=1$ while $F_{v}^{(k)}$ with $k>1$ can satisfy the regularity condition only of $\ell=0$, not one of $\ell=1$. So we cannot impose the regularity condition with $\ell=0$ or $\ell=1$ independently. We should consider these modes at the same time. This means that the perturbations with $\ell=0$ and $\ell=1$ are coupled even at the large $D$ limit. This is the remarkable feature of $\ell=0$ and $\ell=1$ mode perturbations contrasted with $\ell>1$. Then the perturbation equations at $O(1/n^{3/2})$ gives a condition of the frequency for $\ell=0, 1$ modes by
%
\begin{eqnarray}
\hat{\omega}(4m_{\Phi}-4m^{3}_{\Phi}-2R\hat{\omega}+6Rm^{2}_{\Phi}\hat{\omega} -4R^{2}m_{\Phi}\hat{\omega}^{2}+R^{3}\hat{\omega}^{3} )=0,
\label{QNMLO}
\end{eqnarray}
%
where $\hat{\omega}$ is defined by
%
\begin{eqnarray}
\omega = \frac{\sqrt{R^{2}-1}}{R}\frac{\hat{\omega}}{\sqrt{n}}.
\end{eqnarray}
%
The condition (\ref{QNMLO}) has four roots. The four roots can be written by two pairs as
%
\begin{eqnarray}
\hat{\omega} =\frac{m_{\Phi}\pm i\sqrt{m_{\Phi}^{2}-2}}{R},~~ \frac{m_{\Phi}}{R} \mp \frac{m_{\Phi}}{R}.
\label{QNMsmLO}
\end{eqnarray}
%
These two pairs  can be identified with $\ell=0$ and $\ell=1$ modes of the boosted black string respectively by taking the large radius limit $R\rightarrow\infty$ with $m_{\Phi}/R$ fixed\footnote{This fixing is needed since the string direction is identified with $R\Phi$ at the large radius limit.}. As for the perturbations of the boosted black string, see Appendix \ref{B}. So perturbations with $\ell=0$ and $\ell=1$ are coupled at finite ring radius by the condition (\ref{QNMLO}). At the leading order we have only one unstable mode which is the GL like instability of black rings and two marginally stable modes. Going to higher order corrections, we find that another unstable mode, that is, the elastic instability, appears from one of marginally stable modes. Hence the elastically unstable mode couples with the GL like unstable mode by the condition (\ref{QNMLO}). The fact that the GL like instability appears together with the elastic instability has been also observed in numerical analysis \cite{Figueras:2015hkb}. We can obtain the following quasinormal mode formula for $\ell=0$ and $\ell=1$ perturbations up to $O(1/n^{3/2})$ corrections \footnote{
Note that we can obtain the quasinormal mode formula up to $O(1/n^{3/2})$ corrections for black rings by using the effective equations up to $O(1/n^{5/2})$ corrections.
} as
%
\begin{eqnarray}
&&
\omega^{(\ell=0)}_{\pm}=\frac{1}{\sqrt{n}}\frac{\sqrt{R^{2}-1}}{R}\Biggl[
\frac{m_{\Phi}}{R}\pm \frac{i\sqrt{\mP^{2}-2}}{R} \notag \\
&&~~~~
+\frac{1}{\sqrt{n}} \Biggl(\frac{4i\mP^{2}(\mP^{2}-2)(R^{2}-1)-5R^{2}+4}{4(\mP^{2}-1)R^{2}(R^{2}-1)}\pm
\frac{\mP\sqrt{\mP^{2}-2}(4\mP^{2}(R^{2}-1)-9R^{2}+8)}{2(\mP^{2}-1)(\mP^{2}-2)R^{2}(R^{2}-1)} \Biggr) \notag \\
&&~~~~
+\frac{1}{n}\Biggl(
\frac{\mP}{4(\mP^{2}-1)^{3}R^{3}(R^{2}-1)^{2}}\Bigl(
8-18R^{2}+3R^{4}-2R^{6} +2\mP^{6}(R^{2}-4)(R^{2}-1)^{2} \notag \\
&&~~~~~~~~~
+\mP^{4}(40-88R^{2}+45R^{4}-6R^{6})
+\mP^{2}(-72+164R^{2}-81R^{4}+6R^{6}) \notag \\
&&~~~~~~~~~
+2(\mP^{2}-1)^{2}(-2+4R^{2} -R^{4} +2\mP^{2}(R^{2}-1)^{2})\log{(R-R^{-1})}
\Bigr) \notag \\
&&~~~~~~
\pm\frac{i}{32(\mP^{2}-2)^{3/2}(\mP^{2}-1)^{3}R^{3}(R^{2}-1)^{2}}\Bigl(
16\mP^{10}(R^{2}-1)^{2}(3R^{2}+1) \notag \\
&&~~~~~~~~
-16\mP^{8}(6+7R^{2}-30R^{4}+17R^{6}) -2(144-264R^{2}+89R^{4}+32R^{6}) \notag \\
&&~~~~~~~~
+\mP^{2}(1552-2984R^{2} +1127R^{4} +320R^{6}) +\mP^{6}(480-264R^{2} -814R^{4} +592R^{6}) \notag \\
&&~~~~~~~~
-\mP^{4}(1408-2208R^{2} +173R^{4} +624R^{6}) \notag \\
&&~~~~~~~~
-32(\mP^{2}-2)^{2}(\mP^{2}-1)^{3}(R^{2}-1)^{2}\log{(R-R^{-1})}
\Bigr)
\Biggr)
\Biggr]+O(n^{-2}),
\label{QNMBRell0}
\end{eqnarray}
%
and 
%
\begin{eqnarray}
&&
\omega^{(\ell=1)}_{\pm}=\frac{1}{\sqrt{n}}\frac{\sqrt{R^{2}-1}}{R}\Biggl[
\frac{\mP}{R}\mp \frac{\mP}{R} \notag \\
&&~~~~~
+\frac{1}{\sqrt{n}}\Biggl(
\frac{i\mP^{2}(8-9R^{2})}{4(\mP^{2}-1)R^{2}(R^{2}-1)}\mp \frac{i\mP^{2}(4-5R^{2})}{2(\mP^{2}-1)R^{2}(R^{2}-1)}
\Biggr) \notag \\
&&~~~~~
+\frac{1}{n}\Biggl(
\frac{\mP}{4(\mP^{2}-1)^{3}R^{3}(R^{2}-1)^{2}}\Bigl(
-2\mP^{6}R^{2}(4-7R^{2}+3R^{4}) +R^{2}(8-5R^{2}+6R^{4}) \notag \\
&&~~~~~~~~
+\mP^{2}(48-134R^{2}+87R^{4}-18R^{6})
+\mP^{4}(-16+58R^{2} -51R^{4} +18R^{6}) \notag \\
&&~~~~~~~~
+2(\mP^{2}-1)^{2}(-2+4R^{2}-3R^{4} +2\mP^{2}(R^{2}-1)^{2})\log{(R-R^{-1})}
\Bigr) \notag \\
&&~~~~~~
\pm
\frac{\mP}{32(\mP^{2}-1)^{3}R^{3}(R^{2}-1)^{2}}\Bigl(
48(\mP^{8}-R^{2}+1)(R^{2}-1)^{2}  \notag \\
&&~~~~~~~~
-8\mP^{6}(24-55R^{2} +37R^{4} -6R^{6})
+\mP^{4}(416-1008R^{2}+733R^{4}-144R^{6}) \notag \\
&&~~~~~~~~
+\mP^{2}(-576 +1416R^{2} -991R^{4} +144R^{6}) \notag \\
&&~~~~~~~~ 
-32(\mP^{2}-1)^{3}(R^{2}-1)^{2}\log{(R-R^{-1})}
\Bigr)
\Biggr)
\Biggr] 
+O(n^{-2}).
\label{QNMBRell1}
\end{eqnarray}
%
Here we assigned $\ell=0$ and $\ell=1$ to frequencies by identifications with the perturbation of the boosted black string at large radius limit (see Appendix \ref{B}). $\omega^{(\ell=0)}_{+}$ and $\omega^{(\ell=1)}_{+}$ are unstable modes. $\omega^{(\ell=0)}_{+}$ corresponds to the GL like instability, and $\omega^{(\ell=1)}_{+}$ describes the elastic instability of black rings. $\omega_{+}^{(\ell=1)}$ is given explicitly by
%
\begin{eqnarray}
\omega^{(\ell=1)}_{+}=\frac{1}{n}\frac{i\mP^{2}}{4(\mP^{2}-1)R\sqrt{R^{2}-1}}+O(n^{-3/2}).
\label{ef}
\end{eqnarray}
%
Note that the quasinormal mode formula (\ref{QNMBRell0}) and (\ref{QNMBRell1}) break down for $\mP=1$. The reason for this breakdown is not clear at the moment, so we consider the perturbations with $\mP> 1$. The elastic instability has been observed also in the mode with $\mP =1$ numerically \cite{Figueras:2015hkb}. So it would be interesting to clarify why our formula at large $D$ (\ref{ef}) breaks down at $\mP = 1$ and study the elastic instability of that mode at large $D$.

At large radius limit with fixed $\mP/R$, each frequency for unstable modes becomes 
%
\begin{eqnarray}
\omega^{(\ell=0)}_{+}=\frac{\hat{m}_{\Phi}+i\hat{m}_{\Phi}}{\sqrt{n}}-\frac{i\hat{m}_{\Phi}^{2}}{n}+\frac{\hat{m}_{\Phi}-3i\hat{m}_{\Phi}}{2n^{3/2}}+O(1/n^{2}),
\label{LRLell0}
\end{eqnarray}
%
and
%
\begin{eqnarray}
\omega^{(\ell=1)}_{+}=\frac{3\hat{m}_{\Phi}^{3}}{2n\sqrt{n}}+O(1/n^{2}),
\label{LRLell1}
\end{eqnarray}
%
where $\hat{m}_{\Phi}=m_{\Phi}/R$. We can see that the elastic instability disappear with $O(1/R^{2})$, while the GL like instability still exists at the large radius limit. This is consistent with the fact the (boosted) black string has an instability only in S-mode ($\ell=0$ mode) perturbations \cite{Kudoh:2006bp}. The elastic instability occurs due to the finiteness of the ring radius. So it is peculiar feature to the black ring. The decaying rate of the elastic perturbations of the boosted black string is very small with $O(1/n^{2})$ (see Appendix \ref{B}), which is not covered in the formula (\ref{QNMBRell1}). From this we can guess that there is a critical ring radius where the elastic instability appears. The critical ring radius $R_{c}$ of the elastic instability can be estimated as $R_{c}^{2}=O(n)$, and such value of the ring radius is beyond the validity of current setup. So we cannot give concrete values of the critical ring radius, but it should be very large, not $O(1)$, at large $n$.  

The quasinormal mode frequencies at the large radius limit are reproduced by one of the boosted black string with $k=\hat{m}_{\Phi}$ and the boost parameter $\alpha$ given by
%
\begin{eqnarray}
\alpha=1-\frac{1}{2n}+O(1/n^{2}),
\end{eqnarray}
%
(See appendix \ref{B} for the definition of $\alpha$). This value of the boost parameter coincides with the result by the blackfold analysis \cite{Emparan:2007wm} where the boost parameter $\alpha_{\text{bf}}$ is given by\footnote{
The horizon  angular velocity vector $\Omega_{\tH}\pP$ of the $D=n+4$ dimensional black ring is given in the blackfold approach by \cite{Emparan:2007wm}
%
\begin{eqnarray}
\Omega_{\tH}\frac{\partial}{\partial\Phi}=\frac{1}{\sqrt{n+1}}\frac{1}{R} \frac{\partial}{\partial\Phi}= \sqrt{\frac{n}{n+1}}\frac{1}{R} \frac{\partial}{\partial\phi}
\equiv \alpha_{\text{bf}}\frac{1}{R} \frac{\partial}{\partial\phi}, \notag
\end{eqnarray}
%
where we used eq. (\ref{Phirel}).}
%
\begin{eqnarray}
\alpha_{\text{bf}}=\sqrt{\frac{n}{n+1}}.
\end{eqnarray}
%
%
\begin{figure}[t]
 \begin{center}
  \includegraphics[width=65mm,angle=0]{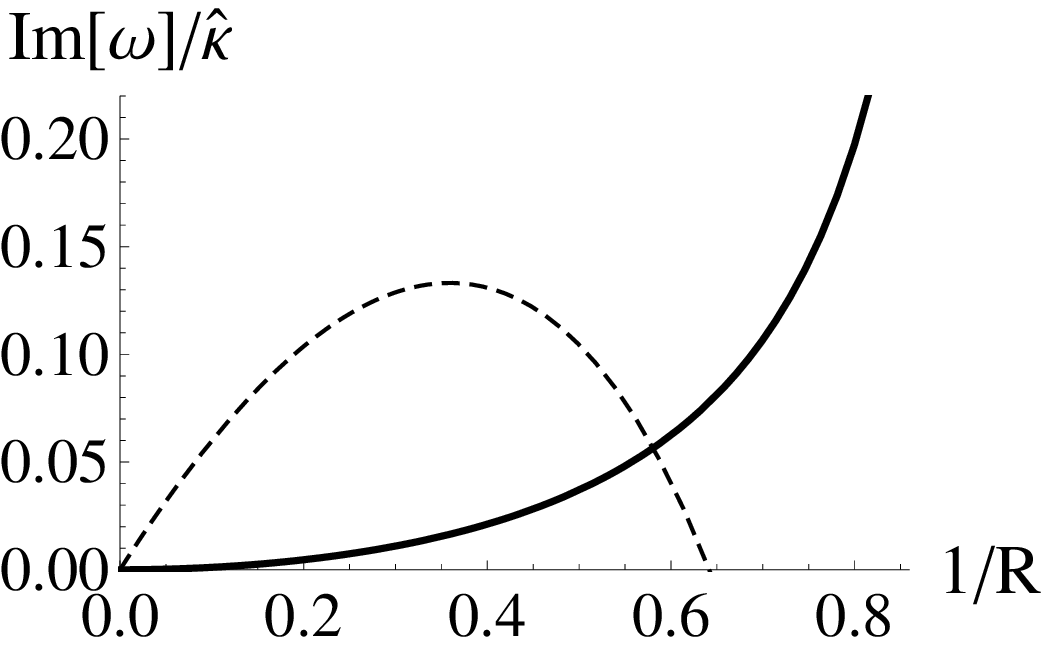}
  \hspace{5mm}
  \includegraphics[width=65mm,angle=0]{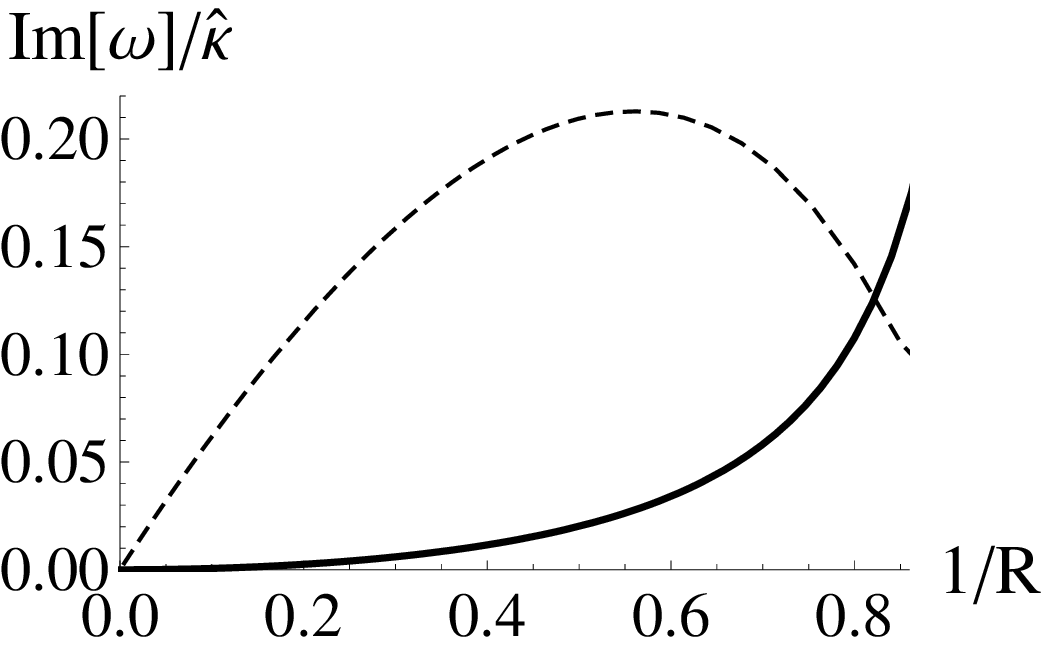}
 \end{center}
 \vspace{-5mm}
 \caption{
Plots of imaginary parts of quasinormal mode frequencies of unstable modes with $\mP=2$. The elastic instability ($\omega^{(\ell=1)}_{+}$) is given by thick line, and the GL like instability ($\omega^{(\ell=0)}_{+}$) by dashed line. The left panel shows the plots for $D=10\,(n=6)$ and right panel is for $D=15\,(n=11)$. As we can see, the GL like instability becomes more dominant in higher dimensions. }
\label{QNM}
\end{figure}
%
In numerical nonlinear analysis on dynamical evolutions of black rings \cite{Figueras:2015hkb}, the elastic instability is dominant dynamics of certain thin black ring in five dimensions. For the perturbation with $m_{\Phi}=O(1)$, the growth rate of the GL like instability is $O(1/\sqrt{n})$, and $O(1/n)$ for the elastic instability at large $n$ in our results. This suggests that the GL like instability becomes dominant over the elastic instability in enough higher dimensions also for certain thin black ring. This behavior can be seen in Figure.\ref{QNM}. In this plot we show imaginary parts of the quasinormal mode frequency of $\omega^{(\ell=1)}_{+}$ (elastic instability) and $\omega^{(\ell=0)}_{+}$ (GL like instability) by thick and dashed lines respectively. The left panel is for $D=10\,(n=6)$, and right panel is for $D=15\,(n=11)$. The surface gravity, $\hat{\kappa}$, was given in eq. (\ref{kappa}). Then, as we can see in plots, the GL like unstable mode becomes more dominant in the parameter region in higher dimensions. These plots well reproduce the behavior of numerical results in \cite{Figueras:2015hkb}.

\section{Summary}
\label{4}

In this paper we studied the elastic instability of the black ring by using the large $D$ expansion method. The different point from our previous analysis on large $D$ black ring \cite{Tanabe:2015hda} is that we should perform $1/\sqrt{D}$, not $1/D$, expansion of the Einstein equations to capture the elastic instability of $D$ dimensional black rings. We found the quasinormal mode formula describing the GL like instability and elastic instability by the perturbation analysis of the effective equations with $1/\sqrt{D}$ expansions. We could identify the original modes of these instabilities in boosted black string perturbations. This large $D$ result for quasinormal mode formula suggests that the elastic instability becomes less efficient than the GL like instability even for certain thin black rings in enough higher dimensions. Our effective equations describe the non-linear evolution of instabilities, so it would be interesting to solve the effective equations by numerical method to study nonlinear evolutions of instabilities. Furthermore the elastic instability is the phenomena of $O(r^{2}_{0}/R^{2})$ where $R$ and $r_{0}$ is the ring radius and horizon thickness respectively. This implies that the elastic instability is triggered by the higher order corrections due to bending effects of the black string horizon in the context of the blackfold approach \cite{Emparan:2007wm, Emparan:2009cs}. Then, as done for the Gregory-Laflamme instability of black branes in \cite{Camps:2010br}, it would be also interesting to investigate the elastic instability using the blackfold approach by including higher order corrections with $O(r_{0}^{2}/R^{2})$ as performed in \cite{Armas:2014bia} to understand general properties of the elastic instability.

\section*{Acknowledgments}
The author is very grateful to Pau Figueras and Roberto Emparan for useful discussions and valuable comments on the draft. 
This work was supported by JSPS Grant-in-Aid for Scientific Research No.26-3387.

\appendix

\section{Ring coordinate embedding}\label{A}

The ring coordinate for $D=n+4$ dimensional flat spacetime is \cite{Emparan:2006mm}
%
\begin{eqnarray}
ds^{2}=-dt^{2} +\frac{R^{2}}{(R+r\cos{\theta})^{2}}\Biggl[
\frac{R^{2}dr^{2}}{R^{2}-r^{2}}
 + (R^{2}-r^{2})d\Phi^{2}
+r^{2}(d\theta^{2} +\sin^{2}{\theta}d\Omega^{2}_{n})
\Biggr],
\label{ringm}
\end{eqnarray}
%
where $0\leq r\leq R$, $0\leq\theta\leq \pi$ and $0\leq\Phi\leq 2\pi$. $R$ is a ring radius. $\theta=0\,(\theta=\pi)$ is an inner (outer) equatorial plane. $r=\text{constant}$ surfaces have $S^{1}\times S^{n+1}$ topology. The black ring solution can be constructed by embedding the leading order solution into a flat spacetime with this ring coordinate. The leading order solution of slowly rotating large $D$ black holes is given by
%
\begin{eqnarray}
&&
ds^{2} = -\left(1-\frac{p_{v}(v,x^{a})}{\sR} \right)dv^{2} +\frac{2H(z)}{\sqrt{1-H'(z)^{2}}}dvdr  \notag \\
&&~~~~~~~~~~~~~~
-\frac{2p_{a}(v,x^{a})}{n\sR}dvdx^{a}
 +G(z)^{2}d\Phi^{2} +dz^{2} +r^{2}H(z)^{2}d\Omega^{2}_{n},
\end{eqnarray}
%
where
%
\begin{eqnarray}
r&=&r_{0}\sR^{1/n} \notag \\
&=&r_{0}+O(1/n).
\end{eqnarray}
%
At large $\sR$ the leading solution approaches
%
\begin{eqnarray}
&&
ds^{2} = -dv^{2} +\frac{2H(z)}{\sqrt{1-H'(z)^{2}}}dvdr  +G(z)^{2}d\Phi^{2}+dz^{2} +r^{2}H(z)^{2}d\Omega^{2}_{n}+O(\sR^{-1}).
\end{eqnarray}
%
Then $r=r_{0}$ surface of the leading order solution is embedded into a flat spacetime in the ring coordinate (\ref{ringm}) by identification $t=v$ with
%
\begin{eqnarray}
H(z)=\frac{R  \sin{\theta}}{R+\cos{\theta}},~~
G(z)=\frac{R \sqrt{R^{2}-1}}{R+\cos{\theta}},~~
\frac{d\theta}{dz}=\frac{R+\cos{\theta}}{R }.
\label{BR2}
\end{eqnarray}
%
Here we set to $r_{0}=1$. In this scale the ring radius $R$ should be larger than unity. This embedding gives the slowly rotating large $D$ black holes with $S^{1}\times S^{n+1}$ horizon topology, which is the black ring in $D=n+4$ dimensions.

\section{Boosted black string}\label{B}

In this appendix we give large $D$ analysis on boosted black strings. This analysis is useful to understand the results of black rings. 

\subsection{Effective equations} 

The leading order solution of slowly rotating large $D$ black hole (\ref{LOsol}) describe dynamical boosted black string by embedding the solution into a $D=n+4$ dimensional flat spacetime with one flat direction whose metric is
%
\begin{eqnarray}
ds^{2}=-dt^{2}+dr^{2}+d\Phi^{2}+r^{2}(d\theta^{2}+\sin^{2}{\theta}d\Omega_{n}^{2}).
\end{eqnarray}
%
$\Phi$ is the flat direction here. The embedding in this coordinate is given by
%
\begin{eqnarray}
G(z)=1,~~H(z)=\sin{z},
\end{eqnarray}
%
in the leading order solution (\ref{LOsol}). The slow rotation, $g_{v\Phi}=O(1/\sqrt{n})$, implies that the boost velocity along $\Phi$ direction is $O(1/\sqrt{n})$. So the solution describes the slowly boosted black string. The effective equations (\ref{Deq1}), (\ref{Deq2}) and (\ref{Deq3}) in this embedding becomes
%
\begin{eqnarray}
\pv p_{v}-\cot{z}~\pz p_{v}+p_{z}\cot{z}+\frac{1}{\sqrt{n}}\pP p_{\phi}=O(1/n),
\end{eqnarray}
%
%
\begin{eqnarray}
\pv p_{\phi}-\cot{z}~\pz p_{\phi}-\cot{z}\frac{p_{z}p_{\phi}}{p_{v}}-\frac{1}{\sqrt{n}}\Biggl[ \pP p_{v} -\pP\Biggl( \frac{p_{\phi}^{2}}{p_{v}} \Biggr) \Biggr] =O(1/n),
\end{eqnarray}
%
and
%
\begin{eqnarray}
\pv p_{z}-\cot{z}~\pz p_{z}+\pz p_{v}  -\cot{z}\frac{p_{z}^{2}}{p_{v}} 
-\frac{\cos{2{z}}}{\sin^{2}{z}}p_{z}+\frac{1}{\sqrt{n}}\partial_{\Phi}\Biggl[ \frac{p_{z}p_{\phi}}{p_{v}} \Biggr] = O(1/n).
\end{eqnarray}
%
We give effective equations up to $O(1/\sqrt{n})$ for simplicity again.  Here we used the relation 
%
\begin{eqnarray}
\phi=\sqrt{n}\Phi. 
\end{eqnarray}
%
In the previous analysis on the GL instability by using the large $D$ expansion method \cite{Asnin:2007rw, Emparan:2013moa, Emparan:2015rva, Suzuki:2015axa, Emparan:2015gva} we assumed $\pP=O(\sqrt{n})$. However, instead, we consider $\pP=O(1)$ case to match with the elastic instability of black rings in this paper.

\subsection{Quasinormal modes}

The effective equations have a stationary solution
%
\begin{eqnarray}
p_{v}(v,z,\Phi)=1,~~p_{z}(v,z,\Phi)=0,~~p_{\phi}(v,z,\Phi)=\alpha,
\end{eqnarray}
%
where $\alpha$ is a boost parameter. This is the $D=n+4$ dimensional boosted black string with $O(1/\sqrt{n})$ boost velocity. We consider the perturbations around this solution as
%
\begin{eqnarray}
p_{v}(v,z,\Phi)=1+\epsilon F_{v}(z)e^{-i\omega v+ik\Phi},\\
p_{z}(v,z,\Phi)=\epsilon F_{z}(z)e^{-i\omega v+ik\Phi},\\
p_{\phi}(v,z,\Phi)=\alpha(1+\epsilon F_{\phi}(z)e^{-i\omega v+ik\Phi}),
\end{eqnarray}
%
where $\epsilon$ is an expansion parameter. We impose the boundary condition at $z=0$ by
%
\begin{eqnarray}
F_{v}(z)\propto z^{\ell}(1+O(z))
\end{eqnarray}
%
with a positive integer. In this appendix we give results only for $\ell=0$ and $\ell=1$. The quasinormal mode frequency is obtained as
%
\begin{eqnarray}
\omega^{(\ell=1)}_{\pm}=\frac{(\mp 1+\alpha)k}{\sqrt{n}}\pm\frac{k(1\mp 2\alpha + 2\alpha^{2}+ 3k^{2})}{2n{\sqrt{n}}}-\frac{3ik^{4}}{n^{2}} +O(1/n^{5/2}),
\label{BBSell1}
\end{eqnarray}
%
for $\ell=1$ and 
%
\begin{eqnarray}
&&
\omega^{(\ell=0)}_{\pm}=\frac{(\pm i +\alpha)k}{\sqrt{n}} -\frac{ik^{2}}{n} \mp\frac{(i \mp 2\alpha +2i\alpha^{2})k}{2n\sqrt{n}}\mp\frac{(2i\mp 6i\alpha+3\alpha^{2})k^{2}}{2n^{2}} \notag \\
&&~~~~~~
\pm\frac{(3i \mp 8\alpha-4i\alpha^{2}\mp 8\alpha^{3} +8ik^{2} \pm 16\alpha k^{2})k}{8n^{2}\sqrt{n}}+O(1/n^{3}),
\label{BBSell0}
\end{eqnarray}
%
for $\ell=0$. For each $\ell$ mode we can study perturbations independently contrasted with the black ring case, where we cannot treat $\ell=0$ and $\ell=1$ modes separately. In the papers \cite{Emparan:2013moa, Emparan:2015rva} we obtained the quasinormal mode frequency of the black string for $\ell=0$ by assuming $\pP=O(\sqrt{n})$ as
%
\begin{eqnarray}
&&
\omega^{(\ell=0)}_{\pm}(\hat{k})= i(\pm\hat{k}-\hat{k}^{2})-\frac{i\hat{k}}{2n}(\pm1+2\hat{k}\mp2\hat{k}^{2}) \notag \\
&&~~~~~~
+\frac{i\hat{k}}{24n^{2}}(\pm9+24\hat{k}\pm12\hat{k}^{2}\mp8\pi^{2}\hat{k}^{3}\mp12\hat{k}^{4})+O(1/n^{3}),
\label{BSell0}
\end{eqnarray}
%
where $\hat{k}=k/\sqrt{n}$. Note that  quasinormal mode frequency $\omega^{(\ell=0)}_{\pm}(\hat{k})$ for black strings has been obtained up to $O(1/n^{4})$ corrections in \cite{Emparan:2015rva}. $\omega^{(\ell=0)}_{\pm}$ for zero boost velocity, $\alpha=0$, in eq. (\ref{BBSell0}) can be reproduced by using $\hat{k}=k/\sqrt{n}$ to eq. (\ref{BSell0}). 

One of important observations is that quasinormal mode frequencies in eqs. (\ref{BBSell1}) and (\ref{BBSell0}) reproduce one of the black ring at the large radius limit (\ref{LRLell0}) and (\ref{LRLell1}) with the identification
%
\begin{eqnarray}
k=\frac{\mP}{R},~~
\alpha = 1 -\frac{1}{2n}+O(1/n^{2}).
\end{eqnarray}
%
Furthermore we can see that the boosted black string is stable for perturbations with $\ell>0$, and there is only one unstable mode in $\ell=0$ perturbation, which is the Gregory-Laflamme instability. The stability of boosted black strings to elastic perturbations ($\ell=1$ modes) can be observed at $O(1/n^{2})$ for the first time. This means that the decaying rate of elastic perturbations is very small with $O(1/n^{2})$ at large $D$. The stability of the boosted black string for $\ell>0$ modes has been observed by the large $D$ expansion method for $k=O(\sqrt{n})$ in \cite{Tanabe:2015hda}. In this paper we could confirm this stability of the boosted black string also for $k=O(1)$ case. 


\end{document}